\documentclass{article}[12pt]
\usepackage{hyperref}
\usepackage{graphicx,amssymb,amsfonts,amsmath,amssymb,amscd,amstext,mathrsfs}
\usepackage{cleveref}
\usepackage{color}

\def\be{\begin{eqnarray}}
\def\ee{\end{eqnarray}}

\begin{document}
\thispagestyle{empty}

\begin{flushright}
LCTP-19-38
\end{flushright}

\begin{center}
\vspace{1cm} { \LARGE {\bf 
Kerr Black Hole Evaporation and Page Curve
}}
\vspace{12mm}

{Jun Nian$^{1,  2}$}
 
\bigskip
{\it
$^1$ International Centre for Theoretical Physics Asia-Pacific,\\ University of Chinese Academy of Sciences, 100190 Beijing, China\\[.5em]
}

{ \it
$^2$ Leinweber Center for Theoretical Physics,\\
University of Michigan,  Ann Arbor,  MI 48109,  U.S.A.}

\vspace{8mm}

{\tt nianjun@ucas.ac.cn}

\vspace{10mm}

\end{center}

\begin{abstract}
\noindent
We compute the black hole entropy and the entanglement entropy of Hawking radiations due to photons during the evaporation of a 4d asymptotically flat Kerr black hole. The Page curve for the Kerr black hole is obtained in the original way \`a la Page,  and it qualitatively mimics the curve for the Schwarzschild black hole but has some new features.

\end{abstract}

\pagebreak
\setcounter{page}{1}

\tableofcontents

\section{Introduction}

Since the milestone works by Bekenstein \cite{Bekenstein:1973ur} and Hawking  \cite{Hawking:1974sw}, there has been a lot of progress in understanding the various aspects of black hole entropy. For instance, the classical Bekenstein-Hawking entropy can be obtained from the microstates in the D1-D5 brane configurations \cite{Strominger:1996sh}. More recently, the asymptotically AdS black hole entropy can also be computed using the boundary CFT in various dimensions \cite{Benini:2015eyy, Cabo-Bizet:2018ehj, Choi:2018hmj, Benini:2018ywd, Choi:2019miv, Larsen:2019oll, Kantor:2019lfo, Nahmgoong:2019hko, Choi:2019zpz, Nian:2019pxj}.

In order to resolve the long-standing black hole information paradox, different proposals have been made in the literature,  including \cite{Giddings:1992hh,  Hartle:1996rp,  Maldacena:2001kr,  Lunin:2001jy,  Horowitz:2003he,  Mathur:2005zp,  Skenderis:2008qn,  Almheiri:2012rt,  Papadodimas:2012aq,  Almheiri:2013hfa,  Papadodimas:2013wnh,  Papadodimas:2013jku,  Bradler:2013gqa,  Hawking:2016msc} and more recently \cite{Penington:2019npb,  Almheiri:2019psf,  Almheiri:2019hni,  Penington:2019kki,  Almheiri:2019qdq,  Nian:2023xmr}. To test these proposals, a crucial tool is the Page curve, which is the time evolution curve of the black hole entropy and the Hawking radiation entanglement entropy during the evaporation process.

Historically, the Page curve was found by Don Page in the study of the von Neumann entropy or the entanglement entropy of the Hawking radiation during the black hole evaporation \cite{Page:1993wv, Page:2013dx}. If the black hole is initially in a pure state, and the evaporation process is unitary, the von Neumann entropy of the Hawking radiation should initially increase and eventually decrease to zero when the black hole disappears. Using the technique of computing the emission rate for massless particles \cite{Page:1976df, Page:1976ki}, the Page curve for the Schwarzschild black hole has been obtained in \cite{Page:1993wv, Page:2013dx}.

When we generalize the previous computations for the Schwarzschild black hole to the Kerr black hole, the principles remain the same. However, the additional angular momentum complicates the problem because it is time-dependent. Nevertheless, using the technique of \cite{Page:1976df, Page:1976ki, Page:2004xp}, we can solve two coupled ordinary differential equations for the mass $M(t)$ and the angular momentum $J(t)$ numerically. Hence, in this paper, we compute the Page curve for the Kerr black hole similar to the original way \`a la Page. We find some interesting features. First, the Page curve for the Kerr black hole mimics the one for the Schwarzschild black hole because the Page curves for both cases initially increase and eventually decrease to zero. Second, for the Kerr black hole, the mass and the angular momentum reduce to zero almost simultaneously. Third, we can compute the Page curves for different initial values of the angular momentum, and the bigger the initial angular momentum $J$ is, the faster the evaporation takes place. Hence, the Kerr black hole generally evaporates faster than the Schwarzschild black hole with the same mass.

This paper is organized as follows. In Sec.~\ref{sec:Review}, we briefly review the notion of the Page curve and its historical developments. In Sec.~\ref{sec:Analytic}, we illustrate the principle of calculating the Page curve and collect some relevant analytic results to the Kerr black hole. Sec.~\ref{sec:Numerical} is the main result of this paper, in which we present the numerical results of the Parge curves with various choices of parameters. A summary and possible future directions are given in Sec.~\ref{sec:Discussion}.

\section{Review of Black Hole Evaporation and Page Curve}\label{sec:Review}

For a Schwarzschild black hole with the mass $M$, the horizon area in the unit $G_N = 1$ is
\be
  A = 16 \pi M^2\, ,
\ee
while the Hawking temperature is
\be
  T_H = \frac{1}{8 \pi M}\, .
\ee
Applying the Stefan-Boltzmann law, we obtain the power of the Hawking radiation through the horizon:
\be
  L = \frac{\Gamma \gamma}{15360\, \pi M^2}\, ,
\ee
where $\Gamma$ and $\gamma$ denote the greybody factor and the total number of massless degrees of freedom,  respectively.  Hence,  the time evolution of the black hole mass is
\be
  \frac{dM}{dt} = - L\, ,
\ee
which leads to the solution
\be\label{eq:Schwarzschild M(t)}
  M(t) = M_0 \left(1 - \frac{t}{t_L} \right)^\frac{1}{3}\, ,
\ee
with $t_L \equiv 5120\, \pi M_0^3 / \Gamma \gamma$ denoting the Schwarzschild black hole's lifetime.

Based on \eqref{eq:Schwarzschild M(t)}, the time evolution of the black hole entropy is
\be
  S_{\textrm{BH}} (t) = 4 \pi M^2 (t) = 4 \pi M_0^2 \left(1 - \frac{t}{t_L} \right)^\frac{2}{3}\, .
\ee
During the Hawking radiation process, the coarse-grained entropy of the black hole decreases, while the coarse-grained entropy of the Hawking radiation increases. However, the increase of the coarse-grained entropy of the Hawking radiation is greater than the coarse-grained entropy of the black hole, which is characterized by the parameter
\be\label{eq:Def beta}
  \beta \equiv \frac{dS_{\textrm{rad}} / dt}{- dS_{\textrm{BH}} / dt}\, ,
\ee
whose numerical value has been given in \cite{Page:2013dx}.  As a warmup example,  we will use the same value of $\beta$ as in \cite{Page:2013dx} for the Schwarzschild black hole.  Consequently, the Hawking radiation entropy is
\be
  S_{\textrm{rad}} = 4 \pi \beta M_0^2 \left[1 - \left(1 - \frac{t}{t_L} \right)^\frac{2}{3} \right]\, .
\ee
The values of $S_{\textrm{BH}}$ and $S_{\textrm{rad}}$ become the same at the Page time $t_*$ given by
\be
  t_* \equiv t_L \left[1 - \left(\frac{\beta}{1 + \beta} \right)^{3/2} \right]\, .
\ee

\begin{figure}[!htb]
\begin{center}
  \includegraphics[width=0.64\textwidth]{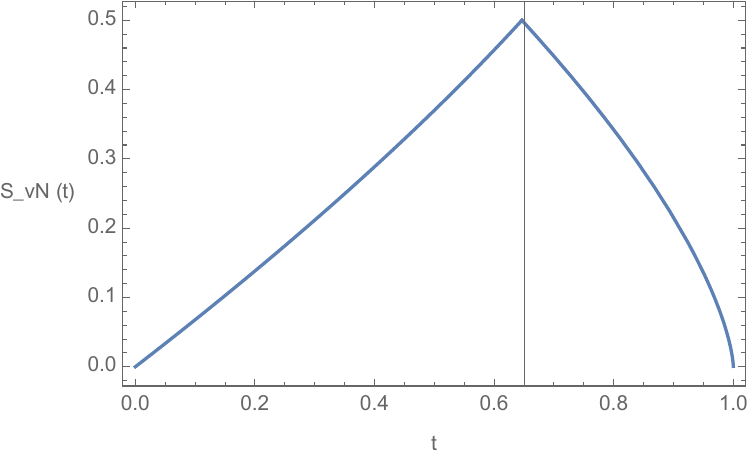}
\end{center}
\vspace{-8mm}
\caption{\small The Page curve for the Schwarzschild black hole}\label{fig:PageCurveSchwarzschild}
\vspace{8mm}
\end{figure}

As shown in \cite{Page:1993wv, Page:2013dx}, the von Neumann entropy or the entanglement entropy of the Hawking radiation, $S_{\textrm{vN}} (t)$, is very close to $S_{\textrm{rad}} (t)$ for $t < t_*$ and very close to $S_{\textrm{BH}}$ for $t > t_*$. Therefore, we can express $S_{\textrm{vN}} (t)$ using the Heaviside step function $\Theta (t)$ as follows:
\begin{align}
  S_{\textrm{vN}} (t) & \approx S_{\textrm{rad}}(t) \cdot \Theta (t_* - t) + S_{\textrm{BH}} (t)\cdot \Theta (t - t_*) \nonumber\\
  {} & = 4 \pi \beta M_0^2 \left[1 - \left(1 - \frac{t}{t_L} \right)^\frac{2}{3} \right]\, \Theta (t_* - t) + 4 \pi M_0^2 \left(1 - \frac{t}{t_L} \right)^\frac{2}{3}\, \Theta (t - t_*)\, ,\label{eq:SvN for Schwarzschild BH}
\end{align}
which is the Page curve for the Schwarzschild black hole (see Fig.~\ref{fig:PageCurveSchwarzschild}).

\vspace{8mm}

\begin{figure}[!htb]
\begin{center}
  \includegraphics[width=0.54\textwidth]{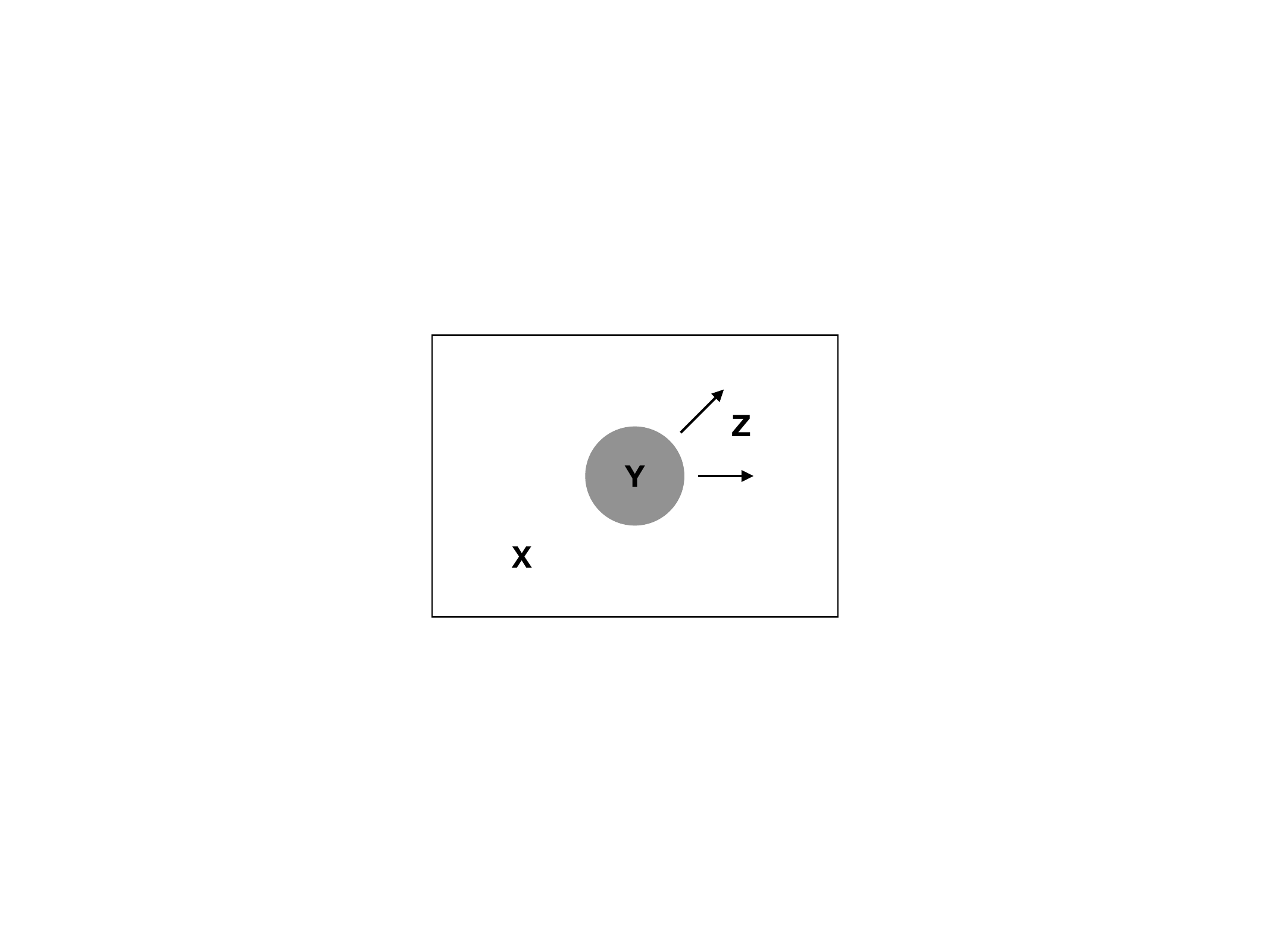}
\end{center}
\vspace{-8mm}
\caption{\small The black hole $Y$ with the reference system $X$ and the Hawking radiation $Z$}\label{fig:BHsystem}
\vspace{8mm}
\end{figure}

When an ambient reference system $X$ is considered, the Page curve of the von Neumann entropy can be distinguished from that of the black hole entropy. As discussed in \cite{Page:2013dx}, we assume that the black hole $Y$ is initially maximally entangled with the reference system $X$, and the Hawking radiation is $Z$ (see Fig.~\ref{fig:BHsystem}).

\begin{figure}[!htb]
\begin{center}
  \includegraphics[width=0.64\textwidth]{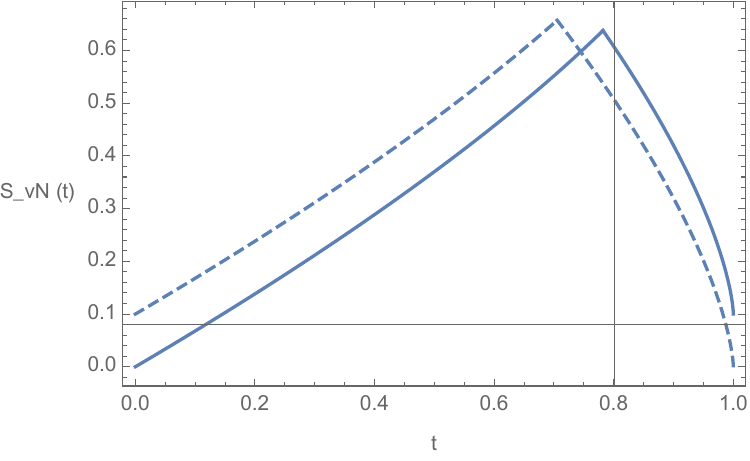}
\end{center}
\vspace{-8mm}
\caption{\small The Page curve for the Schwarzschild black hole with the reference system ($f = 0.1$)}\label{fig:PageCurveSchwarzschildWithRef}
\vspace{8mm}
\end{figure}

Suppose that the reference system has the entropy $S (X) = f\, S_{\textrm{BH}} (0)$ with a constant $0 \leq f \leq 1$. The entropy of the black hole and the one of the Hawking radiation are
\be
  S_Y = S_{\textrm{BH}} (t)\, ,\quad S_Z = \frac{\beta \left[1 - (1 - t / t_L)^{2/3} \right]}{(1 - t / t_L)^{2/3}}\, S_{\textrm{BH}} (t)\, .
\ee
Defining $\tau \equiv 1 - (1 - t / t_L)^{2/3}$, these entropies can be expressed as
\be
  S_X = f\, S_{\textrm{BH}} (0)\, ,\quad S_Y = (1 - \tau)\, S_{\textrm{BH}} (0)\, ,\quad S_Z = \beta \tau S_{\textrm{BH}} (0)\, .
\ee
The time evolution of the Schwarzschild black hole can be divided into three stages:
\begin{enumerate}
\item[(1)] $0 \leq \tau \leq \tau_{12}$: $\quad S_X + S_Z \leq S_Y$.

\item[(2)] $\tau_{12} \leq \tau \leq \tau_{23}$: $\quad S_Y \leq S_X + S_Z \leq 2 S_X + S_Y$.

\item[(3)] $\tau_{23} \leq \tau \leq 1$: $\quad S_X + S_Y \leq S_Z$.
\end{enumerate}
The time scales $\tau_{12}$ and $\tau_{23}$ are defined as
\be
  \tau_{12} \equiv \frac{1 - f}{1 + \beta}\, ,\quad \tau_{23} \equiv \frac{1 + f}{1 + \beta}\, .
\ee
With these refined notions of entropies, the time $t_*$ splits into $t_{12}$ and $t_{23}$. In Fig.~\ref{fig:PageCurveSchwarzschildWithRef}, the von Neumann entropy for the black hole is denoted by the dashed line, while the von Neumann entropy for the Hawking radiation is denoted by the solid line.

\newpage
\section{Analytical Calculation for Kerr Black Hole}\label{sec:Analytic}

The rotating black hole was first constructed by Kerr in \cite{Kerr:1963ud},  called the Kerr black hole in the literature. It is an asymptotically flat 4d black hole with a given angular momentum $J$. In the Boyer-Lindquist coordinates $(t,\, r,\, \theta,\, \phi)$, the Kerr black hole metric can be written as
\begin{align}
  ds^2 & = - \left(1 - \frac{2 M r}{\rho^2} \right) dt^2 + \left(r^2 + a^2 + \frac{2 a^2 M r\, \textrm{sin}^2 \theta}{\rho^2} \right)\, \textrm{sin}^2 \theta\, d\phi^2 - \frac{4 a M r\, \textrm{sin}^2 \theta}{\rho^2}\, d\phi\, dt\nonumber\\
  {} & \quad + \frac{\rho^2}{\Delta}\, dr^2 + \rho^2\, d\theta^2\, , \label{eq:KerrMetric}
\end{align}
where
\be
  \rho^2 = r^2 + a^2\, \textrm{cos}^2 \theta\, ,\quad \Delta \equiv r^2 + a^2 - 2 M r\, ,\quad a \equiv \frac{J}{M}\, .
\ee
The metric \eqref{eq:KerrMetric} describes a 4d non-extremal black hole with the angular momentum $J$. The outer and the inner horizons are
\be
  r_\pm = M \pm \sqrt{M^2 - a^2}\, ,
\ee
and the corresponding Hawking temperature is
\be
  T_H = \frac{r_+ - r_-}{8 \pi M r_+}\, .
\ee
The black hole entropy can be computed using the Bekenstein-Hawking formula:
\be\label{eq:Kerr SBH}
  S_{\textrm{BH}} = 2 \pi M r_+ = 2 \pi M \left(M + \sqrt{M^2 - \frac{J^2}{M^2}} \right)\, .
\ee

The Kerr black hole has played a very important role in theoretical physics. In particular, after the discovery of the AdS/CFT correspondence \cite{Maldacena:1997re, Witten:1998qj}, it was found that the near-horizon region of the extreme Kerr black hole includes a warped AdS$_3$, and consequently the Kerr/CFT correspondence can be established \cite{Guica:2008mu}.  For the non-extremal Kerr black hole, a hidden conformal symmetry was found in \cite{Castro:2010fd}, which recently was used to compute the non-extremal Kerr black hole entropy via the Cardy formula \cite{Haco:2018ske}.

Since the Kerr black hole is characterized by two parameters, the mass $M$ and the angular momentum $J$, in order to obtain the Page curve for the Kerr black hole, we should study the time evolution of both parameters. Using the following notations:
\be
  A = 4 \pi (r_+^2 + a^2)\, ,\qquad \kappa \equiv \frac{4 \pi (r_+ - M)}{A}\, ,\qquad \Omega \equiv \frac{4 \pi a}{A}\, ,
\ee
we can compute the time derivatives of $M(t)$ and $J(t)$ \cite{Page:1976df, Page:1976ki}:
\be\label{eq:ODE}
  - \frac{d}{dt} \left( \begin{array}{c}
  M \\
  J \end{array} \right)
  = \sum_{jlmp} \frac{1}{2 \pi} \int d\omega\, \Gamma_{j \omega l m p} \frac{1}{\textrm{exp} \left[\frac{2 \pi}{\kappa} (\omega - m \Omega) \right] \mp 1}
  \left( \begin{array}{c}
  \omega \\
  m
  \end{array} \right)\, ,
\ee
where $\omega$ denotes the frequency of the $j$-th emission particle, while $l, m, p$ are spherical harmonic quantum number, angular momentum quantum number,  and the polarization of the $j$-th emission particle,  respectively.  The choice of the sign in \eqref{eq:ODE} depends on the spin statistics of the radiated particle species.  The factor $\Gamma_{j \omega l m p}$ can be viewed as a generalized greybody factor, which for the spin-$s$ particle has the following explicit expression:
\be\label{eq:Gamma}
  \Gamma_{s \omega l m p} = \left\{
  \begin{aligned}
    & \left[\frac{(l - s)!\, (l + s)!}{(2 l)!\, (2 l + 1) !!} \right]^2 \prod_{n=1}^l \left[1 + \left(\frac{\omega - m \Omega}{n \kappa} \right)^2 \right] 2 \left(\frac{\omega - m \Omega}{\kappa} \right) \left(\frac{A \kappa}{2 \pi} \omega \right)^{2 l + 1}\, , & \textrm{for } 2s \textrm{ even}\, , \\
    & \left[\frac{(l - s)!\, (l + s)!}{(2 l)!\, (2 l + 1) !!} \right]^2 \prod_{n=1}^{l + 1/2} \left[1 + \left(\frac{\omega - m \Omega}{n \kappa - \frac{1}{2} \kappa} \right)^2 \right] \left(\frac{A \kappa}{2 \pi} \omega \right)^{2 l + 1}\, , & \textrm{for } 2s \textrm{ odd}\, .
  \end{aligned} \right.
\ee
The dominant contribution comes from the $l = s$ modes. In this paper, we only consider the photon emission for simplicity, and we use the leading contribution
\be\label{eq:Gamma spin 1}
  \Gamma_{1 \omega 1 m p} = \frac{4 A}{9 \pi} \Big[M^2 + (m^2 - 1) a^2 \Big] (\omega - m \Omega) \omega^3\quad \textrm{for } s = 1
\ee
to approximate the factor $\Gamma_{j \omega l m p}$.

The Hawking radiation's entropy can be computed directly from the definition of the von Neumann entropy. As discussed in \cite{Page:2004xp}, assuming the absence of the backreaction to geometry, the density matrix of Hawking radiation is the uncorrelated tensor product of thermal density matrices of different modes with definite quantum numbers $(j, \omega, l, m, p)$. The thermal density matrices are diagonal in the number basis, with the probability of $n$ particles in this mode given by
\be
  P_n = N^n\, (1 \pm N)^{- (n \pm 1)}\, ,
\ee
where $N$ denotes the expected number of particles,  $N_{s \omega l m p}$,  related to the factor $\Gamma_{j \omega l m p}$:
\be\label{eq:relation N and Gamma}
  N_{s \omega l m p} = \frac{\Gamma_{s \omega l m p}}{\textrm{exp} \left[\frac{2 \pi}{\kappa} (\omega - m \Omega) \right] \mp 1}\, .
\ee
Hence, the von Neumann entropy of Hawking radiations at a specific mode with fixed quantum numbers $(j, \omega, l, m, p)$ is given by
\be\label{eq:Srad from a single mode}
  S_{\textrm{vN}}^{(j, \omega, l, m, p)} \equiv - P_n\, \textrm{log}\, P_n \Big|_{n=N} = (N \pm 1)\, \textrm{log} (1 \pm N) - N\, \textrm{log}\, N\, ,
\ee
which is a given mode's contribution to the Hawking radiation's von Neumann entropy in unit time. Plugging \eqref{eq:relation N and Gamma} into \eqref{eq:Srad from a single mode} and integrating over all the quantum numbers, we obtain the precise expression for the increasing rate of the Hawking radiation's entropy \cite{Page:1983ug,  Page:2004xp}:
\begin{align}
  \frac{d S_{\textrm{rad}}}{dt} = \sum_{jlmp} \frac{1}{2 \pi} \int d\omega\, & \Bigg[\frac{\Gamma_{j \omega l m p}}{\textrm{exp} \left[\frac{2 \pi}{\kappa} (\omega - m \Omega) \right] \mp 1}\, \textrm{log} \left(\frac{\textrm{exp} \left[\frac{2 \pi}{\kappa} (\omega - m \Omega) \right] \mp 1}{\Gamma_{j \omega l m p}} \pm 1 \right) \nonumber\\
  {} & \pm \textrm{log} \left(1 \pm \frac{\Gamma_{j \omega l m p}}{\textrm{exp} \left[\frac{2 \pi}{\kappa} (\omega - m \Omega) \right] \mp 1} \right) \Bigg]\, ,\label{eq:Srad rate}
\end{align}
where the factor $\Gamma_{j \omega l m p}$ is given by \eqref{eq:Gamma},  or more specifically by \eqref{eq:Gamma spin 1} for photons.

For the computation in this paper, we implicitly use the thermal state to derive the von Neumann entropy as in Hawking's original computation. From this approach, the radiation's von Neumann entropy is monotonically increasing. Applying the same arguments of Page for Schwarzschild black holes, we also approximate the von Neumann entropy of a Kerr black hole by its radiation's entropy in the early stage and by the remaining black hole's entropy at the late stage, separated by the Page time $t_*$. Similar to \eqref{eq:SvN for Schwarzschild BH} for a Schwarzschild black hole, we obtain for a Kerr black hole:
\be\label{eq:SvN for Kerr BH}
  S_{\textrm{vN}} (t) \approx S_{\textrm{rad}} (t) \cdot \Theta (t_* - t) + S_{\textrm{BH}} (t)\cdot \Theta (t - t_*)\, ,
\ee
where the Page time $t_*$ will be determined numerically for given initial conditions.

We will use \eqref{eq:SvN for Kerr BH} to numerically compute the Page curve in the next subsection. The results based on \eqref{eq:SvN for Kerr BH} is just a qualitative result, generalizing the original arguments by Page. In another paper \cite{Nian:2023xmr}, we will discuss how to obtain a Kerr black hole's Page curve more precisely using a microscopic approach, in particular, the decrease of the Page curve in time at late stages of black hole evaporation.

\section{Numerical Results for the Kerr Black Hole Evaporation}\label{sec:Numerical}

By plugging the factor $\Gamma_{1 \omega 1 m p}$ \eqref{eq:Gamma spin 1} into \eqref{eq:ODE}, the integral over $\omega$ can be evaluated exactly using some polylogarithm functions. After summing over the indices, the right-hand side of \eqref{eq:ODE} are functions of $M(t)$ and $J(t)$. Hence, \eqref{eq:ODE} becomes a coupled system of ordinary differential equations, which can be solved numerically. We present some numerical results in this section.

Once we have solved the system of ordinary differential equations \eqref{eq:ODE}, the time evolution of the black hole entropy, $S_{\textrm{BH}} (t)$, can be obtained using \eqref{eq:Kerr SBH}.  For the entropy of the Hawking radiation,  we can apply the same factor $\Gamma_{1 \omega 1 m p}$ and integrate \eqref{eq:Srad rate} to obtain $S_{\textrm{rad}}$ as a function of time directly,  which does not require the parameter $\beta$ \eqref{eq:Def beta} as for the Schwarzschild black hole.

\begin{figure}[!htb]
\centering
\begin{minipage}{.49\textwidth}
  \centering
  \includegraphics[width=0.97\linewidth]{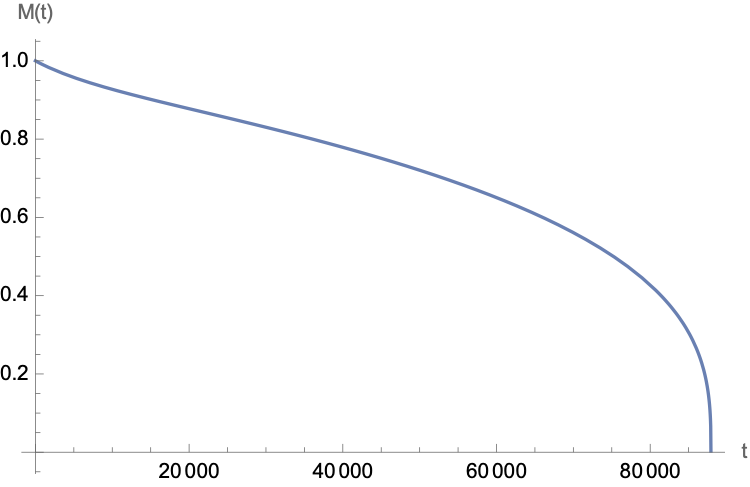}
  \vspace{-4mm}
  \caption{\small The evolution of $M(t)$ for the Kerr black hole with $M(0) = 1$, $J(0) = 1/2$}\label{fig:M(t)}
  \vspace{4mm}
\end{minipage}
\begin{minipage}{.02\textwidth}

\end{minipage}
\begin{minipage}{.49\textwidth}
  \centering
  \includegraphics[width=0.97\linewidth]{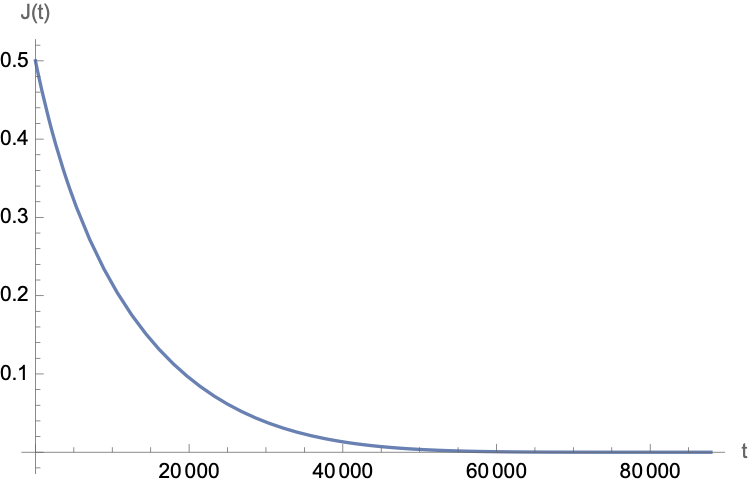}
  \vspace{-4mm}
  \caption{\small The evolution of $J(t)$  for the Kerr black hole with $M(0) = 1$, $J(0) = 1/2$ and $f = 0.2$}\label{fig:J(t)}
  \vspace{4mm}
\end{minipage}
\end{figure}

\begin{figure}[!htb]
\begin{center}
  \includegraphics[width=0.57\textwidth]{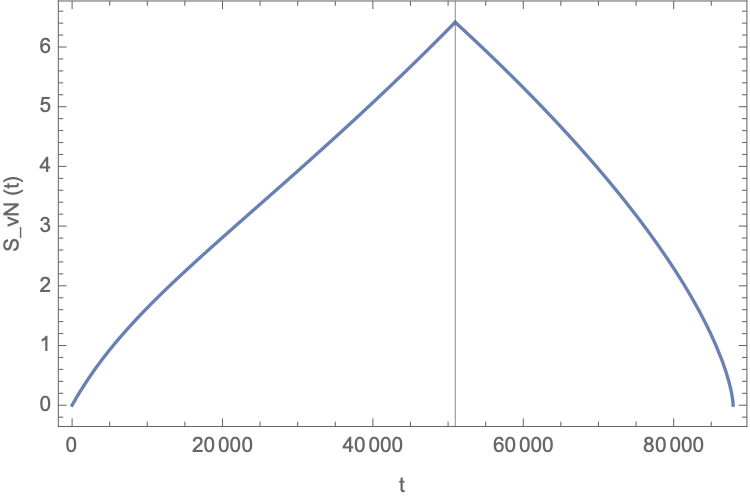}
\end{center}
\vspace{-8mm}
\caption{\small The Page curve for the Kerr black hole with $M(0) = 1$ and $J(0) = 1/2$}\label{fig:Kerr Page Curve J=1/2}
\vspace{8mm}
\end{figure}

Let us first look at an example with the initial conditions $M(0) = 1$ and $J(0) = 1/2$.  The time evolutions of $M(t)$ and $J(t)$ are shown in Figs.~\ref{fig:M(t)} and \ref{fig:J(t)},  while the von Neumann entropy is shown in Fig.~\ref{fig:Kerr Page Curve J=1/2}.

\vspace{8mm}

\begin{figure}[!htb]
\begin{center}
  \includegraphics[width=0.57\textwidth]{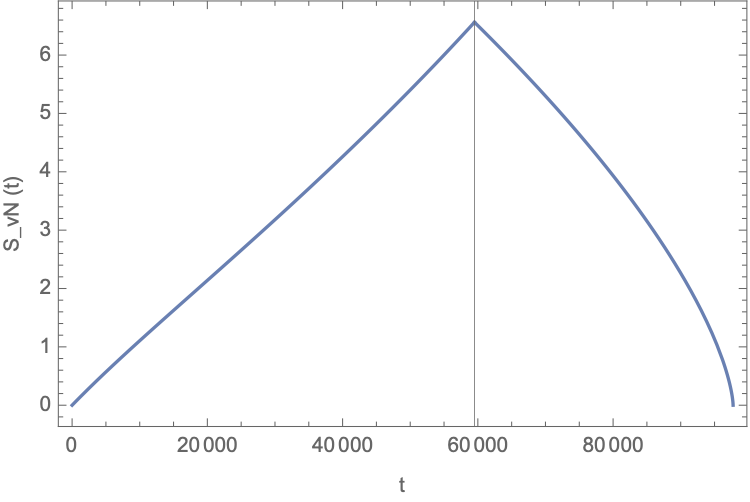}
\end{center}
\vspace{-8mm}
\caption{\small The Page curve for the Kerr black hole with $M(0) = 1$ and $J(0) = 1/4$}\label{fig:Kerr Page Curve J=1/4}
\vspace{8mm}
\end{figure}

\begin{figure}[!htb]
\begin{center}
  \includegraphics[width=0.57\textwidth]{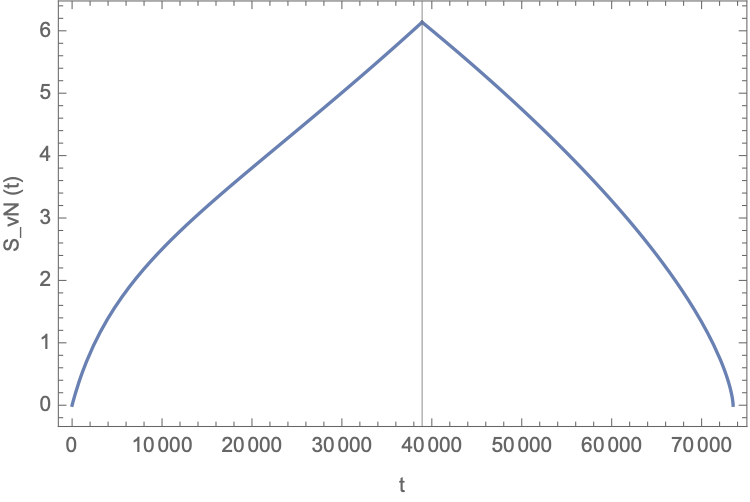}
\end{center}
\vspace{-8mm}
\caption{\small The Page curve for the Kerr black hole with $M(0) = 1$ and $J(0) = 3/4$}\label{fig:Kerr Page Curve J=3/4}
\vspace{8mm}
\end{figure}

Compared to the Schwarzschild black hole, the Kerr black hole evaporation has several important features we want to emphasize.  In general,  both the mass $M(t)$ and the angular momentum $J(t)$ decrease in time.  We observe that $M(t)$ and $J(t)$ vanish almost simultaneously at the end of the black hole evaporation (see Figs.~\ref{fig:M(t)} and \ref{fig:J(t)}),  i.e.,  a Kerr black hole does not evolve into a Schwarzschild black hole during its entire evaporation process.  However,  to the best of my knowledge, no first principles guarantee this fact.  I changed several different initial conditions of $J(0)$,  and this feature holds for all the cases.  Another noticeable feature from Fig.~\ref{fig:Kerr Page Curve J=1/2} through Fig.~\ref{fig:Kerr Page Curve J=3/4} is that,  the larger $J(0)$ is,  the faster the black hole evaporation takes place.  Moreover,  the Page time $t_*$ is not precisely half of a Kerr black hole's lifetime. Instead,  it depends on the initial conditions.  The numerical values of the Page time $t_*$ and the lifetime $t_L$ for different initial conditions are listed in Tab.~\ref{tab:NumValues}.

\begin{table}[htb!]
\centering
\begin{tabular}{|c|c|c|c|}
\hline
$J(0)$ & Page time $t_*$ & Lifetime $t_L$ & $t_* / t_L$ \\
\hline
$1/4$ & 59502.73 & 97747.34 & 0.60874\\
\hline
$1/2$ & 50945.91 & 87903.37 & 0.57957\\
\hline
$3/4$ & 38918.04 & 73502.81 & 0.52948 \\
\hline
\end{tabular}
\caption{The numerical values of $t_*$ and lifetime of a Kerr black hole with $M(0) = 1$ \label{tab:NumValues}}
\end{table}

\begin{figure}[!htb]
\centering
\begin{minipage}{.49\textwidth}
  \centering
  \includegraphics[width=0.97\linewidth]{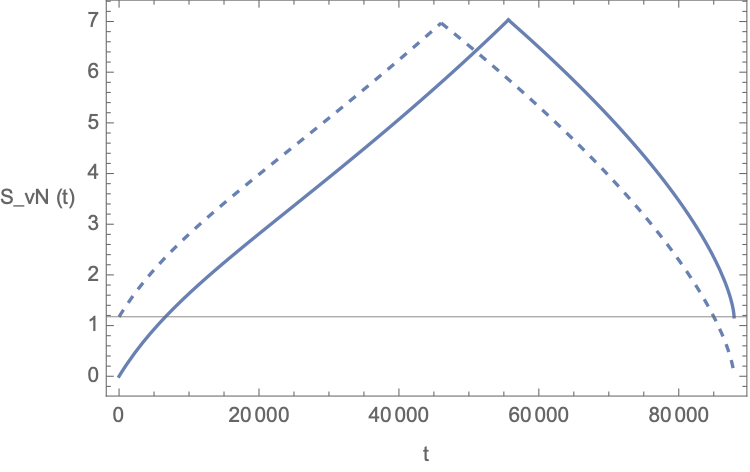}
  \vspace{-4mm}
  \caption{\small The Page curve for the Kerr black hole with $M(0) = 1$, $J(0) = 1/2$ and $f = 0.1$}\label{fig:KerrPageCurveRefined1}
  \vspace{4mm}
\end{minipage}
\begin{minipage}{.02\textwidth}

\end{minipage}
\begin{minipage}{.49\textwidth}
  \centering
  \includegraphics[width=0.97\linewidth]{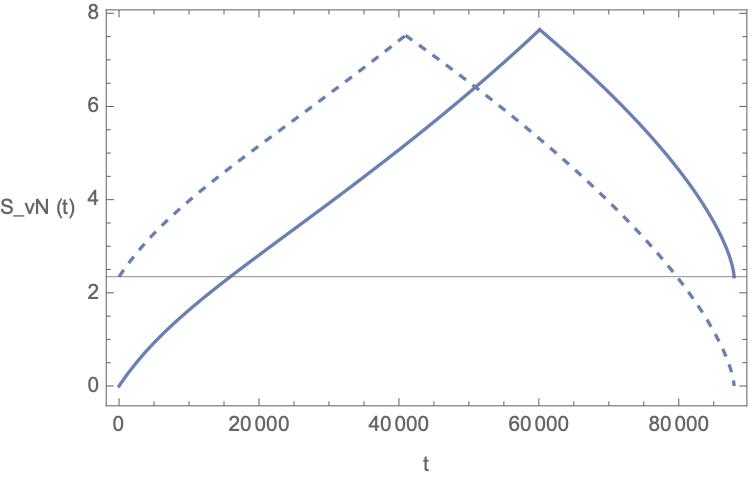}
  \vspace{-4mm}
  \caption{\small The Page curve for the Kerr black hole with $M(0) = 1$, $J(0) = 1/2$ and $f = 0.2$}\label{fig:KerrPageCurveRefined2}
  \vspace{4mm}
\end{minipage}
\end{figure}

\begin{figure}[!htb]
\centering
\begin{minipage}{.49\textwidth}
  \centering
  \includegraphics[width=0.97\linewidth]{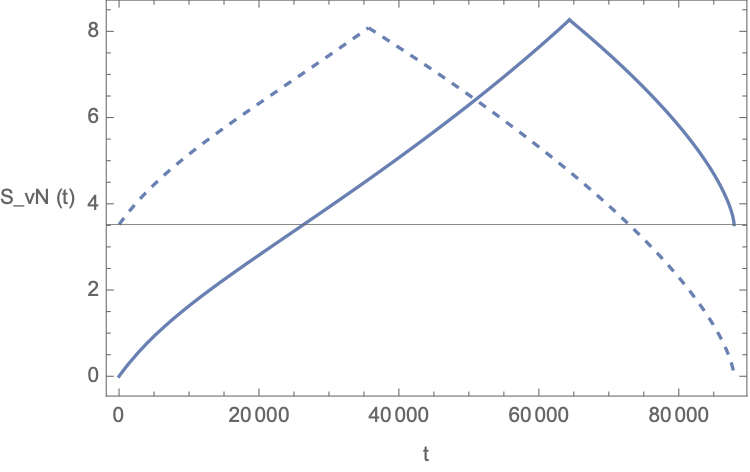}
  \vspace{-4mm}
  \caption{\small The Page curve for the Kerr black hole with $M(0) = 1$, $J(0) = 1/2$ and $f = 0.3$}\label{fig:KerrPageCurveRefined3}
  \vspace{4mm}
\end{minipage}
\begin{minipage}{.02\textwidth}

\end{minipage}
\begin{minipage}{.49\textwidth}
  \centering
  \includegraphics[width=0.97\linewidth]{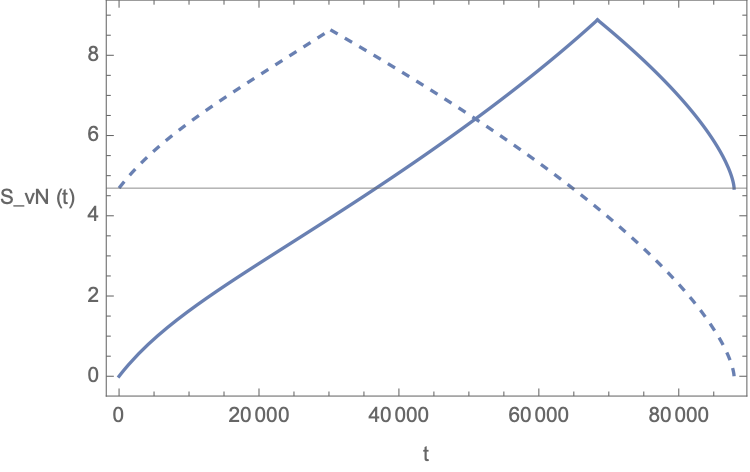}
  \vspace{-4mm}
  \caption{\small The Page curve for the Kerr black hole with $M(0) = 1$, $J(0) = 1/2$ and $f = 0.4$}\label{fig:KerrPageCurveRefined4}
  \vspace{4mm}
\end{minipage}
\end{figure}

\begin{figure}[!htb]
\centering
\begin{minipage}{.49\textwidth}
  \centering
  \includegraphics[width=0.97\linewidth]{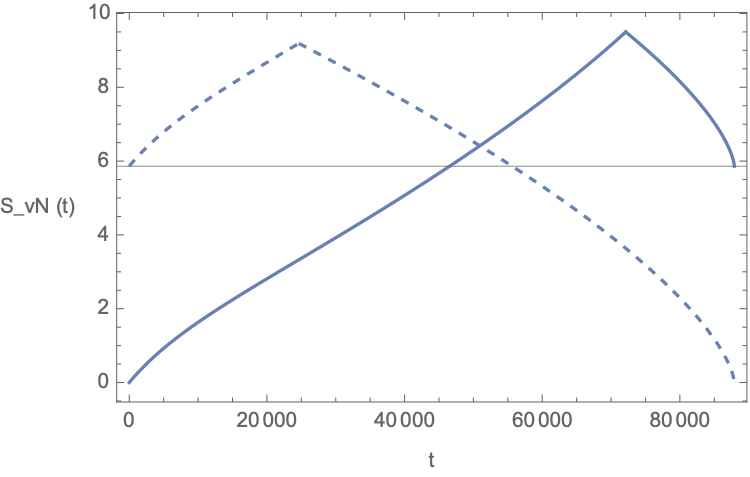}
  \vspace{-4mm}
  \caption{\small The Page curve for the Kerr black hole with $M(0) = 1$, $J(0) = 1/2$ and $f = 0.5$}\label{fig:KerrPageCurveRefined5}
  \vspace{4mm}
\end{minipage}
\begin{minipage}{.02\textwidth}

\end{minipage}
\begin{minipage}{.49\textwidth}
  \centering
  \includegraphics[width=0.97\linewidth]{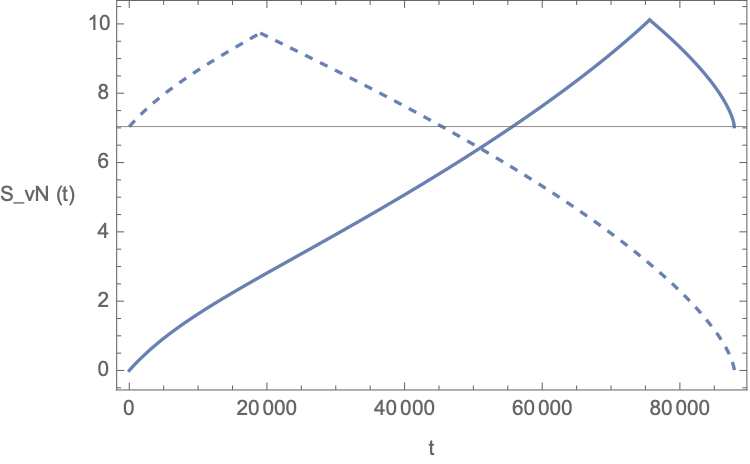}
  \vspace{-4mm}
  \caption{\small The Page curve for the Kerr black hole with $M(0) = 1$, $J(0) = 1/2$ and $f = 0.6$}\label{fig:KerrPageCurveRefined6}
  \vspace{4mm}
\end{minipage}
\end{figure}

\begin{figure}[!htb]
\centering
\begin{minipage}{.49\textwidth}
  \centering
  \includegraphics[width=0.97\linewidth]{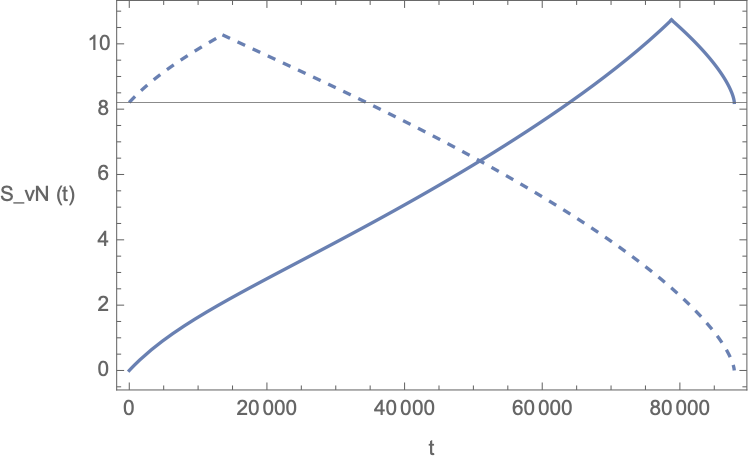}
  \vspace{-4mm}
  \caption{\small The Page curve for the Kerr black hole with $M(0) = 1$, $J(0) = 1/2$ and $f = 0.7$}\label{fig:KerrPageCurveRefined7}
  \vspace{4mm}
\end{minipage}
\begin{minipage}{.02\textwidth}

\end{minipage}
\begin{minipage}{.49\textwidth}
  \centering
  \includegraphics[width=0.97\linewidth]{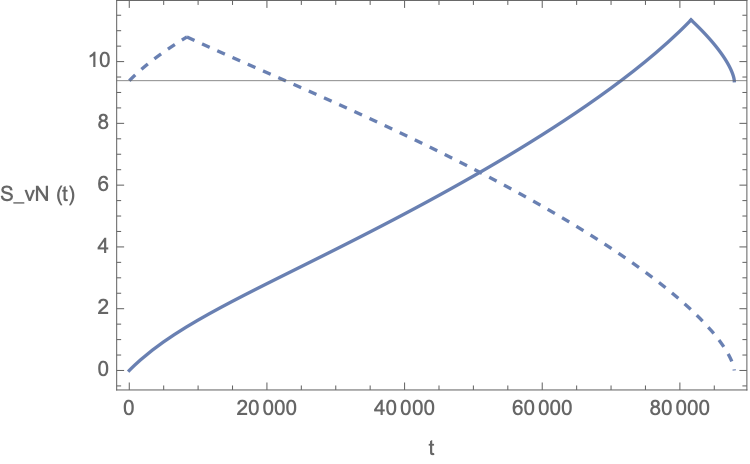}
  \vspace{-4mm}
  \caption{\small The Page curve for the Kerr black hole with $M(0) = 1$, $J(0) = 1/2$ and $f = 0.8$}\label{fig:KerrPageCurveRefined8}
  \vspace{4mm}
\end{minipage}
\end{figure}

\begin{figure}[!htb]
\centering
\begin{minipage}{.49\textwidth}
  \centering
  \includegraphics[width=0.97\linewidth]{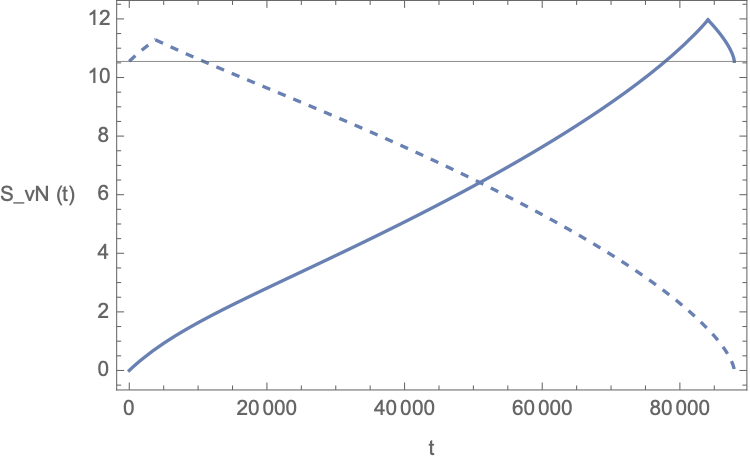}
  \vspace{-4mm}
  \caption{\small The Page curve for the Kerr black hole with $M(0) = 1$, $J(0) = 1/2$ and $f = 0.9$}\label{fig:KerrPageCurveRefined9}
  \vspace{4mm}
\end{minipage}
\begin{minipage}{.02\textwidth}

\end{minipage}
\begin{minipage}{.49\textwidth}
  \centering
  \includegraphics[width=0.97\linewidth]{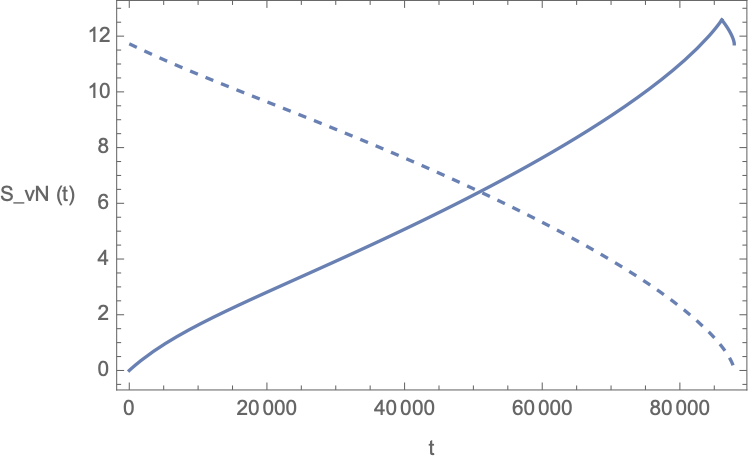}
  \vspace{-4mm}
  \caption{\small The Page curve for the Kerr black hole with $M(0) = 1$, $J(0) = 1/2$ and $f = 1.0$}\label{fig:KerrPageCurveRefined10}
  \vspace{4mm}
\end{minipage}
\end{figure}

\newpage
Similar to the Schwarzschild black hole case discussed in Sec.~\ref{sec:Review}, we can refine the Page curve of the Kerr black hole by considering the black hole initially maximally entangled with a reference system. Then the time $t_*$ splits into $t_{12}$ and $t_{23}$, and the black hole evaporation can be divided into three stages, as described in Sec.~\ref{sec:Review}. We show the numerical results explicitly for the Kerr black hole with $M(0) = 1$ and $J (0) = 1/2$ starting from $f=0.1$ through $f=1$ in Fig.~\ref{fig:KerrPageCurveRefined1} through Fig.~\ref{fig:KerrPageCurveRefined10},  where all the entropies of the Hawking radiation are computed by integrating \eqref{eq:Srad rate} without using the parameter $\beta$.  These refined Page curves show features similar to the Schwarzschild case \cite{Page:2013dx}.

\newpage

\section{Discussion}\label{sec:Discussion}

In this paper, we have computed the Page curve, i.e., the time evolution of the entanglement entropy of the Hawking radiation due to photons for the 4d Kerr black hole. We have seen that the Page curve mimics the one for the Schwarzschild black hole but has some new features. For instance, the mass and the angular momentum of a Kerr black hole decrease to zero almost simultaneously at the end of the evaporation. Moreover, the bigger the initial angular momentum $J$ is, the faster evaporation occurs. Hence, the Kerr black hole generally evaporates faster than the Schwarzschild black hole with the same mass.

Since we have only considered the photon emission in this paper, the natural extension is to incorporate the contributions from other (nearly) massless particles, such as gravitons and neutrinos. The next step is to consider the Kerr-Newman black hole evaporation with a nonzero initial electric charge. This case will undoubtedly have some new features that have not been exposed before because we have to consider the emission of charged particles (e.g., electrons) so that the electric charge of the black hole decreases. For this case, one probably needs to combine the approaches for the Kerr black hole \cite{Page:1976ki} and for the Reissner-Nordstr\"om black hole \cite{Page:1977um} and develop some new techniques. I would like to explore these generalizations in the near future.

This work and the previous results \cite{Page:1993wv, Page:2013dx} provide precise numerical results for the Page curve; hence, they also make a playground for the various proposals for the black hole information paradox \cite{Giddings:1992hh,  Hartle:1996rp,  Maldacena:2001kr,  Lunin:2001jy,  Horowitz:2003he,  Mathur:2005zp,  Skenderis:2008qn,  Almheiri:2012rt,  Papadodimas:2012aq,  Almheiri:2013hfa,  Papadodimas:2013wnh,  Papadodimas:2013jku,  Bradler:2013gqa,  Hawking:2016msc}. The correct model must reproduce the corresponding Page curve quantitatively, which can be used as a guiding principle to test these proposals and eventually help resolve the problem.  The Page curve for an evaporating Kerr black hole has been computed microscopically in \cite{Nian:2023xmr},  and the curve there matches the results in this paper,  supporting the consistency of the microscopic model proposed in \cite{Nian:2023xmr}.

\vskip 0.1in

\section*{Acknowledgments}

I would like to thank the referee for the valuable comments and the critical reading of the manuscript.  I also thank Sung-Soo Kim,  Futoshi Yagi,  Leo A. Pando Zayas,  and Yang Zhou for helpful discussions.  The University of Electronic Science and Technology of China,  Southwest Jiaotong University,  and Sichuan University provided very warm hospitality during the final stage of this work.  I want to thank these institutes.  This work was supported in part by the NSFC under grant No.~12375067,  the U.S. Department of Energy under grant DE-SC0007859, and the Van Loo Postdoctoral Fellowship.

\appendix

\bibliographystyle{utphys}
\bibliography{KerrBHEva}

\providecommand{\href}[2]{#2}\begingroup\raggedright\begin{thebibliography}{10}

\bibitem{Bekenstein:1973ur}
J.~D. Bekenstein, ``{Black holes and entropy},''
\href{http://dx.doi.org/10.1103/PhysRevD.7.2333}{{\em Phys. Rev.} {\bfseries
  D7} (1973) 2333--2346}.

\bibitem{Hawking:1974sw}
S.~W. Hawking, ``{Particle Creation by Black Holes},''
  \href{http://dx.doi.org/10.1007/BF02345020, 10.1007/BF01608497}{{\em Commun.
  Math. Phys.} {\bfseries 43} (1975) 199--220}.
[,167(1975)].

\bibitem{Strominger:1996sh}
A.~Strominger and C.~Vafa, ``{Microscopic origin of the Bekenstein-Hawking
  entropy},'' \href{http://dx.doi.org/10.1016/0370-2693(96)00345-0}{{\em Phys.
  Lett.} {\bfseries B379} (1996) 99--104},
\href{http://arxiv.org/abs/hep-th/9601029}{{\ttfamily arXiv:hep-th/9601029
  [hep-th]}}.

\bibitem{Benini:2015eyy}
F.~Benini, K.~Hristov, and A.~Zaffaroni, ``{Black hole microstates in AdS$_{4}$
  from supersymmetric localization},''
  \href{http://dx.doi.org/10.1007/JHEP05(2016)054}{{\em JHEP} {\bfseries 05}
  (2016) 054},
\href{http://arxiv.org/abs/1511.04085}{{\ttfamily arXiv:1511.04085 [hep-th]}}.

\bibitem{Cabo-Bizet:2018ehj}
A.~Cabo-Bizet, D.~Cassani, D.~Martelli, and S.~Murthy, ``{Microscopic origin of
  the Bekenstein-Hawking entropy of supersymmetric AdS$_{\bf 5}$ black
  holes},''
\href{http://arxiv.org/abs/1810.11442}{{\ttfamily arXiv:1810.11442 [hep-th]}}.

\bibitem{Choi:2018hmj}
S.~Choi, J.~Kim, S.~Kim, and J.~Nahmgoong, ``{Large AdS black holes from
  QFT},''
\href{http://arxiv.org/abs/1810.12067}{{\ttfamily arXiv:1810.12067 [hep-th]}}.

\bibitem{Benini:2018ywd}
F.~Benini and P.~Milan, ``{Black holes in 4d $\mathcal{N}=4$
  Super-Yang-Mills},''
\href{http://arxiv.org/abs/1812.09613}{{\ttfamily arXiv:1812.09613 [hep-th]}}.

\bibitem{Choi:2019miv}
S.~Choi and S.~Kim, ``{Large AdS$_6$ black holes from CFT$_5$},''
\href{http://arxiv.org/abs/1904.01164}{{\ttfamily arXiv:1904.01164 [hep-th]}}.

\bibitem{Larsen:2019oll}
F.~Larsen, J.~Nian, and Y.~Zeng, ``{AdS$_5$ Black Hole Entropy near the BPS
  Limit},''
\href{http://arxiv.org/abs/1907.02505}{{\ttfamily arXiv:1907.02505 [hep-th]}}.

\bibitem{Kantor:2019lfo}
G.~K\'antor, C.~Papageorgakis, and P.~Richmond, ``{AdS$_7$ Black-Hole Entropy
  and 5D $\mathcal{N}=2$ Yang-Mills},''
\href{http://arxiv.org/abs/1907.02923}{{\ttfamily arXiv:1907.02923 [hep-th]}}.

\bibitem{Nahmgoong:2019hko}
J.~Nahmgoong, ``{6d superconformal Cardy formulas},''
\href{http://arxiv.org/abs/1907.12582}{{\ttfamily arXiv:1907.12582 [hep-th]}}.

\bibitem{Choi:2019zpz}
S.~Choi, C.~Hwang, and S.~Kim, ``{Quantum vortices, M2-branes and black
  holes},''
\href{http://arxiv.org/abs/1908.02470}{{\ttfamily arXiv:1908.02470 [hep-th]}}.

\bibitem{Nian:2019pxj}
J.~Nian and L.~A. Pando~Zayas, ``{Microscopic Entropy of Rotating Electrically
  Charged AdS$_4$ Black Holes from Field Theory Localization},''
\href{http://arxiv.org/abs/1909.07943}{{\ttfamily arXiv:1909.07943 [hep-th]}}.

\bibitem{Giddings:1992hh}
S.~B. Giddings, ``{Black holes and massive remnants},''
  \href{http://dx.doi.org/10.1103/PhysRevD.46.1347}{{\em Phys. Rev. D}
  {\bfseries 46} (1992) 1347--1352},
  \href{http://arxiv.org/abs/hep-th/9203059}{{\ttfamily arXiv:hep-th/9203059}}.

\bibitem{Hartle:1996rp}
J.~B. Hartle, ``{Generalized quantum theory in evaporating black hole
  space-times},'' in {\em {Symposium on Black Holes and Relativistic Stars
  (dedicated to memory of S. Chandrasekhar)}}, pp.~195--219.
\newblock 12, 1996.
\newblock \href{http://arxiv.org/abs/gr-qc/9705022}{{\ttfamily
  arXiv:gr-qc/9705022}}.

\bibitem{Maldacena:2001kr}
J.~M. Maldacena, ``{Eternal black holes in anti-de Sitter},''
  \href{http://dx.doi.org/10.1088/1126-6708/2003/04/021}{{\em JHEP} {\bfseries
  04} (2003) 021}, \href{http://arxiv.org/abs/hep-th/0106112}{{\ttfamily
  arXiv:hep-th/0106112}}.

\bibitem{Lunin:2001jy}
O.~Lunin and S.~D. Mathur, ``{AdS / CFT duality and the black hole information
  paradox},'' \href{http://dx.doi.org/10.1016/S0550-3213(01)00620-4}{{\em Nucl.
  Phys. B} {\bfseries 623} (2002) 342--394},
  \href{http://arxiv.org/abs/hep-th/0109154}{{\ttfamily arXiv:hep-th/0109154}}.

\bibitem{Horowitz:2003he}
G.~T. Horowitz and J.~M. Maldacena, ``{The Black hole final state},''
  \href{http://dx.doi.org/10.1088/1126-6708/2004/02/008}{{\em JHEP} {\bfseries
  02} (2004) 008}, \href{http://arxiv.org/abs/hep-th/0310281}{{\ttfamily
  arXiv:hep-th/0310281}}.

\bibitem{Mathur:2005zp}
S.~D. Mathur, ``{The Fuzzball proposal for black holes: An Elementary
  review},'' \href{http://dx.doi.org/10.1002/prop.200410203}{{\em Fortsch.
  Phys.} {\bfseries 53} (2005) 793--827},
  \href{http://arxiv.org/abs/hep-th/0502050}{{\ttfamily arXiv:hep-th/0502050}}.

\bibitem{Skenderis:2008qn}
K.~Skenderis and M.~Taylor, ``{The fuzzball proposal for black holes},''
  \href{http://dx.doi.org/10.1016/j.physrep.2008.08.001}{{\em Phys. Rept.}
  {\bfseries 467} (2008) 117--171},
  \href{http://arxiv.org/abs/0804.0552}{{\ttfamily arXiv:0804.0552 [hep-th]}}.

\bibitem{Almheiri:2012rt}
A.~Almheiri, D.~Marolf, J.~Polchinski, and J.~Sully, ``{Black Holes:
  Complementarity or Firewalls?},''
  \href{http://dx.doi.org/10.1007/JHEP02(2013)062}{{\em JHEP} {\bfseries 02}
  (2013) 062}, \href{http://arxiv.org/abs/1207.3123}{{\ttfamily arXiv:1207.3123
  [hep-th]}}.

\bibitem{Papadodimas:2012aq}
K.~Papadodimas and S.~Raju, ``{An Infalling Observer in AdS/CFT},''
  \href{http://dx.doi.org/10.1007/JHEP10(2013)212}{{\em JHEP} {\bfseries 10}
  (2013) 212},
\href{http://arxiv.org/abs/1211.6767}{{\ttfamily arXiv:1211.6767 [hep-th]}}.

\bibitem{Almheiri:2013hfa}
A.~Almheiri, D.~Marolf, J.~Polchinski, D.~Stanford, and J.~Sully, ``{An
  Apologia for Firewalls},''
  \href{http://dx.doi.org/10.1007/JHEP09(2013)018}{{\em JHEP} {\bfseries 09}
  (2013) 018}, \href{http://arxiv.org/abs/1304.6483}{{\ttfamily arXiv:1304.6483
  [hep-th]}}.

\bibitem{Papadodimas:2013wnh}
K.~Papadodimas and S.~Raju, ``{Black Hole Interior in the Holographic
  Correspondence and the Information Paradox},''
  \href{http://dx.doi.org/10.1103/PhysRevLett.112.051301}{{\em Phys. Rev.
  Lett.} {\bfseries 112} no.~5, (2014) 051301},
  \href{http://arxiv.org/abs/1310.6334}{{\ttfamily arXiv:1310.6334 [hep-th]}}.

\bibitem{Papadodimas:2013jku}
K.~Papadodimas and S.~Raju, ``{State-Dependent Bulk-Boundary Maps and Black
  Hole Complementarity},''
  \href{http://dx.doi.org/10.1103/PhysRevD.89.086010}{{\em Phys. Rev. D}
  {\bfseries 89} no.~8, (2014) 086010},
  \href{http://arxiv.org/abs/1310.6335}{{\ttfamily arXiv:1310.6335 [hep-th]}}.

\bibitem{Bradler:2013gqa}
K.~Br\'adler and C.~Adami, ``{The capacity of black holes to transmit quantum
  information},'' \href{http://dx.doi.org/10.1007/JHEP05(2014)095}{{\em JHEP}
  {\bfseries 05} (2014) 095}, \href{http://arxiv.org/abs/1310.7914}{{\ttfamily
  arXiv:1310.7914 [quant-ph]}}.

\bibitem{Hawking:2016msc}
S.~W. Hawking, M.~J. Perry, and A.~Strominger, ``{Soft Hair on Black Holes},''
  \href{http://dx.doi.org/10.1103/PhysRevLett.116.231301}{{\em Phys. Rev.
  Lett.} {\bfseries 116} no.~23, (2016) 231301},
  \href{http://arxiv.org/abs/1601.00921}{{\ttfamily arXiv:1601.00921
  [hep-th]}}.

\bibitem{Penington:2019npb}
G.~Penington, ``{Entanglement Wedge Reconstruction and the Information
  Paradox},''
\href{http://arxiv.org/abs/1905.08255}{{\ttfamily arXiv:1905.08255 [hep-th]}}.

\bibitem{Almheiri:2019psf}
A.~Almheiri, N.~Engelhardt, D.~Marolf, and H.~Maxfield, ``{The entropy of bulk
  quantum fields and the entanglement wedge of an evaporating black hole},''
\href{http://arxiv.org/abs/1905.08762}{{\ttfamily arXiv:1905.08762 [hep-th]}}.

\bibitem{Almheiri:2019hni}
A.~Almheiri, R.~Mahajan, J.~Maldacena, and Y.~Zhao, ``{The Page curve of
  Hawking radiation from semiclassical geometry},''
\href{http://arxiv.org/abs/1908.10996}{{\ttfamily arXiv:1908.10996 [hep-th]}}.

\bibitem{Penington:2019kki}
G.~Penington, S.~H. Shenker, D.~Stanford, and Z.~Yang, ``{Replica wormholes and
  the black hole interior},''
  \href{http://dx.doi.org/10.1007/JHEP03(2022)205}{{\em JHEP} {\bfseries 03}
  (2022) 205}, \href{http://arxiv.org/abs/1911.11977}{{\ttfamily
  arXiv:1911.11977 [hep-th]}}.

\bibitem{Almheiri:2019qdq}
A.~Almheiri, T.~Hartman, J.~Maldacena, E.~Shaghoulian, and A.~Tajdini,
  ``{Replica Wormholes and the Entropy of Hawking Radiation},''
  \href{http://dx.doi.org/10.1007/JHEP05(2020)013}{{\em JHEP} {\bfseries 05}
  (2020) 013}, \href{http://arxiv.org/abs/1911.12333}{{\ttfamily
  arXiv:1911.12333 [hep-th]}}.

\bibitem{Nian:2023xmr}
J.~Nian, ``{Hawking Radiation, Entanglement Entropy, and Information Paradox of
  Kerr Black Holes},'' \href{http://arxiv.org/abs/2312.14287}{{\ttfamily
  arXiv:2312.14287 [hep-th]}}.

\bibitem{Page:1993wv}
D.~N. Page, ``{Information in black hole radiation},''
  \href{http://dx.doi.org/10.1103/PhysRevLett.71.3743}{{\em Phys. Rev. Lett.}
  {\bfseries 71} (1993) 3743--3746},
\href{http://arxiv.org/abs/hep-th/9306083}{{\ttfamily arXiv:hep-th/9306083
  [hep-th]}}.

\bibitem{Page:2013dx}
D.~N. Page, ``{Time Dependence of Hawking Radiation Entropy},''
  \href{http://dx.doi.org/10.1088/1475-7516/2013/09/028}{{\em JCAP} {\bfseries
  1309} (2013) 028},
\href{http://arxiv.org/abs/1301.4995}{{\ttfamily arXiv:1301.4995 [hep-th]}}.

\bibitem{Page:1976df}
D.~N. Page, ``{Particle Emission Rates from a Black Hole: Massless Particles
  from an Uncharged, Nonrotating Hole},''
\href{http://dx.doi.org/10.1103/PhysRevD.13.198}{{\em Phys. Rev.} {\bfseries
  D13} (1976) 198--206}.

\bibitem{Page:1976ki}
D.~N. Page, ``{Particle Emission Rates from a Black Hole. 2. Massless Particles
  from a Rotating Hole},''
\href{http://dx.doi.org/10.1103/PhysRevD.14.3260}{{\em Phys. Rev.} {\bfseries
  D14} (1976) 3260--3273}.

\bibitem{Page:2004xp}
D.~N. Page, ``{Hawking radiation and black hole thermodynamics},''
  \href{http://dx.doi.org/10.1088/1367-2630/7/1/203}{{\em New J. Phys.}
  {\bfseries 7} (2005) 203},
\href{http://arxiv.org/abs/hep-th/0409024}{{\ttfamily arXiv:hep-th/0409024
  [hep-th]}}.

\bibitem{Kerr:1963ud}
R.~P. Kerr, ``{Gravitational field of a spinning mass as an example of
  algebraically special metrics},''
\href{http://dx.doi.org/10.1103/PhysRevLett.11.237}{{\em Phys. Rev. Lett.}
  {\bfseries 11} (1963) 237--238}.

\bibitem{Maldacena:1997re}
J.~M. Maldacena, ``{The Large N limit of superconformal field theories and
  supergravity},'' \href{http://dx.doi.org/10.1023/A:1026654312961,
  10.4310/ATMP.1998.v2.n2.a1}{{\em Int. J. Theor. Phys.} {\bfseries 38} (1999)
  1113--1133}, \href{http://arxiv.org/abs/hep-th/9711200}{{\ttfamily
  arXiv:hep-th/9711200 [hep-th]}}.
[Adv. Theor. Math. Phys.2,231(1998)].

\bibitem{Witten:1998qj}
E.~Witten, ``{Anti-de Sitter space and holography},''
  \href{http://dx.doi.org/10.4310/ATMP.1998.v2.n2.a2}{{\em Adv. Theor. Math.
  Phys.} {\bfseries 2} (1998) 253--291},
\href{http://arxiv.org/abs/hep-th/9802150}{{\ttfamily arXiv:hep-th/9802150
  [hep-th]}}.

\bibitem{Guica:2008mu}
M.~Guica, T.~Hartman, W.~Song, and A.~Strominger, ``{The Kerr/CFT
  Correspondence},'' \href{http://dx.doi.org/10.1103/PhysRevD.80.124008}{{\em
  Phys. Rev.} {\bfseries D80} (2009) 124008},
\href{http://arxiv.org/abs/0809.4266}{{\ttfamily arXiv:0809.4266 [hep-th]}}.

\bibitem{Castro:2010fd}
A.~Castro, A.~Maloney, and A.~Strominger, ``{Hidden Conformal Symmetry of the
  Kerr Black Hole},'' \href{http://dx.doi.org/10.1103/PhysRevD.82.024008}{{\em
  Phys. Rev.} {\bfseries D82} (2010) 024008},
\href{http://arxiv.org/abs/1004.0996}{{\ttfamily arXiv:1004.0996 [hep-th]}}.

\bibitem{Haco:2018ske}
S.~Haco, S.~W. Hawking, M.~J. Perry, and A.~Strominger, ``{Black Hole Entropy
  and Soft Hair},'' \href{http://dx.doi.org/10.1007/JHEP12(2018)098}{{\em JHEP}
  {\bfseries 12} (2018) 098},
\href{http://arxiv.org/abs/1810.01847}{{\ttfamily arXiv:1810.01847 [hep-th]}}.

\bibitem{Page:1983ug}
D.~N. Page, ``{COMMENT ON `ENTROPY EVAPORATED BY A BLACK HOLE'},''
  \href{http://dx.doi.org/10.1103/PhysRevLett.50.1013}{{\em Phys. Rev. Lett.}
  {\bfseries 50} (1983) 1013}.

\bibitem{Page:1977um}
D.~N. Page, ``{Particle Emission Rates from a Black Hole. 3. Charged Leptons
  from a Nonrotating Hole},''
\href{http://dx.doi.org/10.1103/PhysRevD.16.2402}{{\em Phys. Rev.} {\bfseries
  D16} (1977) 2402--2411}.

\end{thebibliography}\endgroup

\end{document}